\documentclass[groupedaddress,
 amsmath,amssymb, twocolumn, aps, prl, showpacs, showkeys]{revtex4-1}

\usepackage[latin1]{inputenc}
\usepackage{dcolumn}
\usepackage{bm}
\usepackage[pdftex]{graphicx}
\usepackage{nonfloat}
\usepackage{hyperref}
\usepackage{endnotes}

\usepackage{ulem}
\usepackage{color}

\begin{document}

\title{Enhanced harmonic generation in relativistic laser plasma interaction}
\preprint{APS/surface harmonic generation}

\author{C.~Rödel,$^{1,2}$}
\author{E.~Eckner$^{1}$}
\author{J.~Bierbach$^{1,2}$}
\author{M.~Yeung,$^{3}$}
\author{B.~Dromey$^{3}$}
\author{T.~Hahn$^{4}$}
\author{S.~Fuchs$^{1,2}$}
\author{A.~Galestian$^{1}$}
\author{M.~Wünsche$^{1}$}
\author{S.~Kuschel$^{1,2}$}
\author{D.~Hemmers$^{4}$} 
\author{O.~Jäckel$^{1,2}$}
\author{G.~Pretzler$^{4}$}    
\author{M.~Zepf$^{2,3}$} 
\author{G.~G.~Paulus$^{1,2}$}

\affiliation{$^{1}$Institute of Optics and Quantum Electronics, Friedrich-Schiller-University Jena, Germany}
\affiliation{$^{2}$Helmholtz Institute Jena, Germany}
\affiliation{$^{3}$Centre for Plasma Physics, Queen's University Belfast, United Kingdom}
\affiliation{$^{4}$Institute for Laser and Plasma Physics, University of Düsseldorf, Germany}

\begin{abstract}
We report the enhancement of individual harmonics generated at a relativistic ultra-steep plasma vacuum interface. Simulations show the harmonic emission to be due to the coupled action of two high velocity oscillations -- at the fundamental $\omega_L$ and at the plasma frequency $\omega_P$ of the bulk plasma. The synthesis of the enhanced harmonics can be described by the reflection of the incident laser pulse at a relativistic mirror oscillating at $\omega_L$ and $\omega_P$.
\end{abstract}



\maketitle

Waveform shaping and frequency synthesis based on waveform modulation is ubiquitous in electronics, telecommunication technology, and optics  \cite{Kundu2013, Lee2002}. For optical waveforms, the carrier frequency $\omega_L$ is on the order of several hundred THz, while the modulation frequencies used in conventional devices like electro- or acousto-optical modulators are orders of magnitude lower \cite{Lee2002}. As a consequence, any new frequencies are typically very close to the fundamental. The synthesis of new frequencies in the extreme ultraviolet (XUV), e.g. by using relativistic oscillating mirrors \cite{Lichters1996, Tsakiris2006}, requires modulation frequencies in the optical regime \cite{Dromey2006, Thaury2007} or even in the extreme ultraviolet. The latter has not been proven possible to date.


Phase modulation of light upon reflection from relativistically oscillating plasma surfaces has been proposed as a mechanism for efficient high-harmonic generation (HHG). For reviews see Ref.~\cite{Quere2010} \& \cite{Teubner2009}. In this process, a dense plasma from a solid density surface located at $x(t)$ oscillates such that relativistic velocities $\beta=v/c \lesssim 1$ are reached. At times when the relativistic Lorentz factor $\gamma(t)=1/\sqrt{1-\beta^2}$ is large and the plasma is moving towards the incident light wave with high velocity $\beta^{\rm max}$, high frequency radiation is generated in the reflected beam by the relativistic Doppler effect \cite{Gordienko2004,Quere2010}. Due to the mirror-like reflection with a periodicity of $\omega_L$, this process can be interpreted as a phase modulation \cite{Bulanov1994, Lichters1996, Tsakiris2006}. Consequently, it is often referred to as the Relativistic Oscillating Mirror (ROM). Note that the relativistic plasma oscillation $x(t)=\frac{\beta^{\rm max}_L}{\omega_L} \sin(\omega_L t+\varphi_L)$ discussed in the literature to date is at the laser frequency $\omega_L$ or its low order harmonics $n \omega_L$ \cite{Lichters1996,Tsakiris2006}, i.\,e. at optical frequencies. Models that include the relativistic effect of retardation $t=t'-x(t')/c$ show good agreement with numerical simulations \cite{Lichters1996,Quere2010} by accounting for the difference of the time of reflection $t'$ and the time of observation $t$. In the limit of a strongly relativistic motion, such models predict a harmonic spectral envelope spanning a broad bandwidth in the XUV and soft x-ray range that follows a power law $(\omega/\omega_L)^{-8/3}$ up to some roll-off frequency $\omega_{\rm ro}\propto \gamma^3$ \cite{Baeva2006} after which the harmonic intensity decreases exponentially. Moreover, the shallow spectral slope not only implies that coherent high frequency radiation can be efficiently generated by this harmonic generation mechanism, but rather is also a signature of a train of extremely short optical pulses in the time domain \cite{Baeva2006, Quere2010, Behmke2011}. HHG from relativistic surfaces thus is one of the most promising routes for intense attosecond pulse generation \cite{Tsakiris2006, Sansone2011}.

\begin{figure*}[t]
	\centering
		\includegraphics[width=1.0\textwidth]{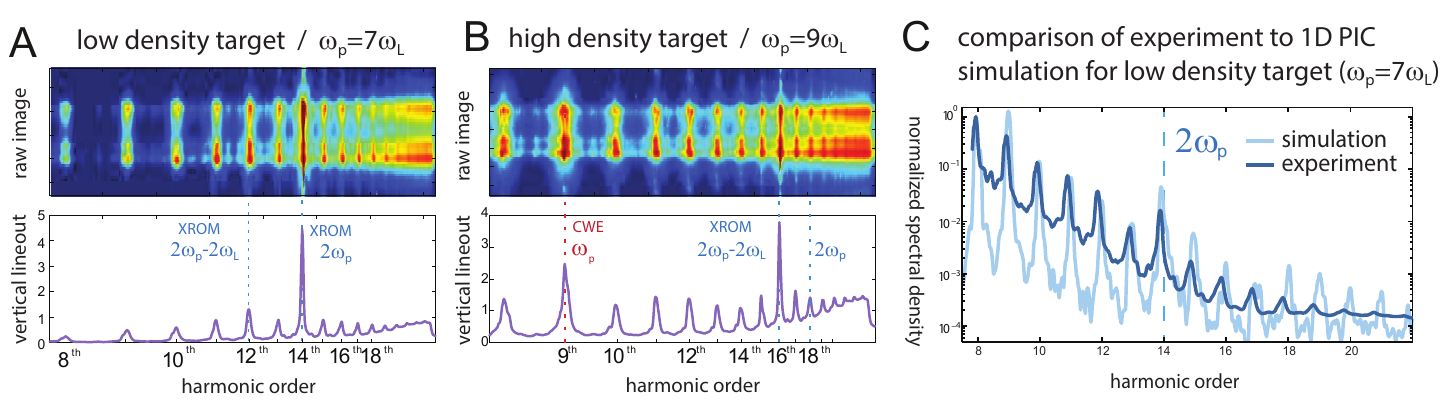}
	\caption{\textbf{High harmonic spectra for different target densities} (upper graph: raw data from CCD camera, the structure in vertical dimenstion is due to aberration of the torodal mirror in the spectrograph; lower graph: vertical lineout) \textbf{A:} The high harmonic spectrum using a plastic coated glass surface as a target material ($\omega_P \simeq 7 \omega_L$) shows a strongly increased $14^{th}$ harmonic at $\simeq 2\omega_P$. \textbf{B:} Using a surface of fused silica the plasma frequency is increased to $\omega_P \approx 9 \omega_L$ (CWE harmonic at $\omega_P$). The enhanced harmonic appears at $2\omega_P-2\omega_L$. \textbf{C:} A laser plasma simulation is performed using the parameters of the experiment with the low density target ($a_0=2.5, \, n_e=49n_c,\, \omega_P=7\omega_L$). The spectral density obtained from the 1D PIC simulation is in good agreement with the experiment. In particular, the enhanced harmonic at $2\omega_P$ is reproduced.}
	\label{spectra}
\end{figure*}

In the last decade, relativistic surface high harmonic generation has been demonstrated up to keV photon energies \cite{Dromey2006} in experiments which have also verified the shallow harmonic slope and the scaling of $\omega_{\rm ro}$ \cite{Dromey2007}. At more moderate intensities of the order of $10^{19} \rm \, W/cm^2$, relativistic surface harmonic radiation has been generated for different laser and plasma parameters \cite{Thaury2007, Tarasevitch2007, Krushelnick2008}. Recently, efficiencies ranging from $10^{-4}$ -- $10^{-6}$ with a strong dependence on the plasma gradient $L_p$ have been reported and generally reveal a decrease of the harmonics' efficiency for very short scale lengths $L_p \ll \lambda/10 $ \cite{Rodel2012, Kahaly2013}. For an $L_p$ of $\lambda/10$ to $\lambda/5$, i.e. for plasma conditions which can be considered to be ideal for efficient generation \cite{Rodel2012, Dollar2013}, a constant, nearly diffraction limited beam divergence has been found for ROM harmonics \cite{Dromey2009}. All experiments so far report a decaying spectral envelope towards higher harmonic orders, possibly with small modulations spanning a few harmonic orders \cite{Watts2002, Teubner2003, Behmke2011} which are explained within the existing theoretical framework of the ROM mechanism where the surface oscillates only at optical frequencies.  

Here we report on the strongly enhanced emission of particular harmonic orders under conditions where the plasma density profile is approaching a step function. Depending on the target density, either the 14th (43.4\,eV), 16th (49.6\,eV), or 18th (55.8\,eV) harmonic of a 400-nm driving terawatt laser pulse has been amplified by a factor of up to five as compared to the adjacent harmonics (see Fig.~\ref{spectra}A/B). The enhancement is never observed for odd harmonic orders or nonrelativistic intensities. It should be noted that the efficiency of enhanced harmonic generation is remarkably high even though the plasma scale length is very short and thus not optimal for conventional ROM harmonics \cite{note}. In fact, the enhanced harmonics contain a pulse energy of $\approx 1 \mu J$ and thus qualify for applications like coherent diffraction imaging \cite{Sandberg2007} or seeding of free electron lasers \cite{Lambert2008}. 

The experimental setup is identical to the one used in a previous experiment for generating ROM harmonics with 10 Hz \cite{Bierbach2012}: The surface harmonics are generated by focusing 400-nm laser pulses with a pulse duration of 45\,fs and a pulse energy of 100\,mJ onto rotating plastic or glass targets such that a fresh interaction surface is provided for each laser pulse. The 400-nm radiation is created by a table-top 40-TW laser system operating at 800\,nm via second harmonic generation (SHG) in a KDP crystal (0.7\,mm). Using an off-axis parabolic mirror, the 400-nm pulses are focused to intensities of $I=2 \cdot 10^{19}\rm\,W/cm^2$ (FWHM) which correspond to a normalized vector potential $a_0=2$. An angle of incidence of $45^\circ$ and laser polarization parallel to the plane of incidence is used.

\begin{figure*}[t]
	\centering
		\includegraphics[width=1.0\textwidth]{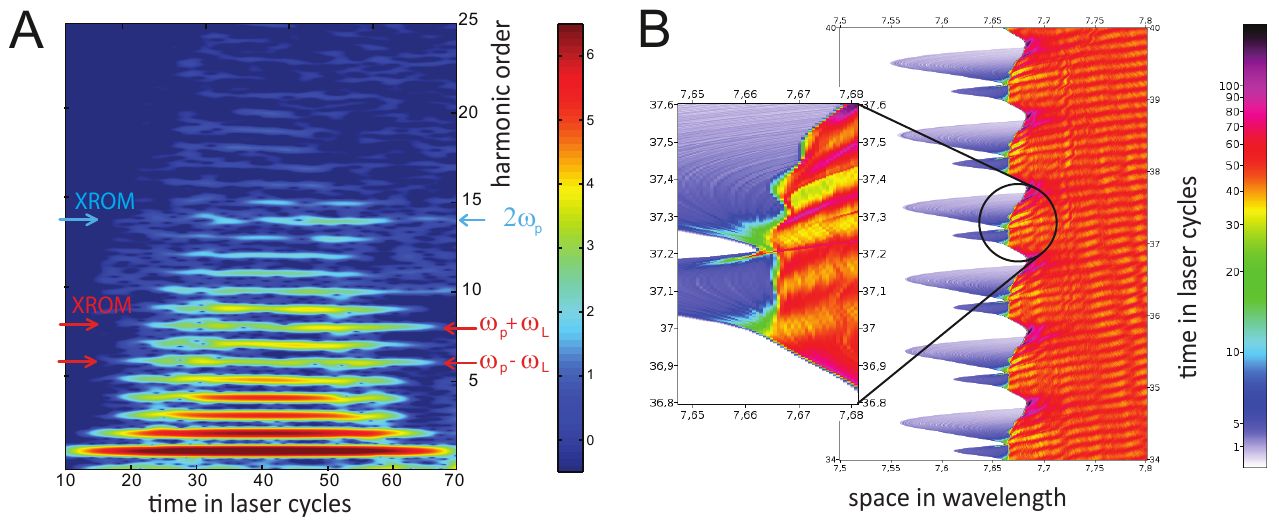}
	\caption{\textbf{Laser plasma simulation A:} The reflected field is analyzed using a time-windowed Fourier transform. The harmonic intensity (logarithmic scale) at $2\omega_P$ is increased as compared to their adjacent harmonics thus reproducing our experimental result. In addition, the XROM harmonic at $2\omega_P$ is emitted for a slightly longer time than the other harmonics in this frequency range - an effect which is also observed at the sideband harmonics of the plasma frequency located at $\omega_P \pm \omega_L$. It is shown that these harmonics are also increased by the XROM. 
	  \textbf{B:} The electron density as a function of time is displayed showing that the plasma is oscillating with different frequencies. Most prominently, the surface plasma  is oscillating with $\omega_L$ and, in addition, with the plasma frequency $\omega_P$ (zoom in). Jets of Brunel electrons (blue) are excited mainly with $\omega_L$ into the dense plasma where they excite Langmuir waves at $\omega_P$.}
	\label{PIC}
\end{figure*}

In the following we show that the enhanced harmonics are generated by a relativistic plasma surface oscillating at XUV frequencies. Therefore, we refer to the high-frequency (XUV) ROM as XROM. A first indication of the origin of the enhanced harmonics is found in the observation that fluctuations in the intensity of the 400-nm fundamental do \textit{not} affect the order of the enhanced harmonic. However, the order of the enhanced harmonic is affected by the target material. For plastic targets (low density), it is the $14^{th}$ harmonic (28.6\,nm), for glass targets (high density) either the $16^{th}$ or $18^{th}$. The enhanced harmonics are in fact conspicuously close to twice the maximum plasma frequency $\omega_P=\sqrt{n_e e^2/(\varepsilon_0 m_e)}$ which is the natural frequency for electrons in a plasma with an electron density $n_e$. Here, $e$ is the charge of an electron, $\varepsilon_0$ the vacuum permittivity and $m_e$ the electron mass. The value of $\omega_P$ was verified to be near the $7^{th}$ harmonic for plastic targets or near the $9^{th}$ harmonic for glass targets, respectively, by measuring the cutoff frequencies for harmonics at lower, non-relativistic intensities. At nonrelativistic conditions it is well established that Coherent Wake Emission (CWE) \cite{Quere2006} becomes dominant and that the high harmonic spectrum ends at $\omega_P$ \cite{Thaury2007}. It is worth noting that the CWE is strongest for the cutoff harmonic located at $\omega_P$, e.g. the $9^{th}$ harmonic in Fig.~\ref{spectra}B. This is a signature of an extremely steep plasma density ramp \cite{Dromey2009a}. However, in this frequency range the complex superposition of harmonics generated by different generation mechanisms makes it nearly impossible to distinguish between CWE and ROM harmonics, or -- as we will see later -- the XROM process. 

Before explaining the physical details of the XROM, we briefly review known alternative mechanisms for the generation of radiation at the plasma frequency and its multiples. Plasma radiation at $\omega_P$ and $2\omega_P$ is a well-known observation in gas discharges, tokamak plasmas \cite{Fidone1978} and astrophysical plasmas such as auroral \cite{Forsyth1949} and solar radio burst \cite{Wild1953} and is, in general, incoherent. In laser produced plasmas, radiation at $\omega_P$ and $2\omega_P$ \cite{Teubner1999} has been observed and attributed to the coupling of two large-amplitude Langmuir waves together with subsequent two-plasmon decay (TPD) \cite{Boyd2000, Kunzl2003}. Common to this type of incoherent, broadband plasma radiation is the requirement of plasma waves generated inside the plasma and, in general, their nonrelativistic origin. Thus such radiation should not be strongly intensity dependent and their spatial and spectral properties would be unlikely to resemble those of ROM harmonic orders. Conversely, the similarity of ROM- and XROM-harmonics in terms of divergence, intensity dependence and bandwidth indicates that their spatial and temporal phases are compatible to the ROM mechanism and hence suggest strongly that the origin of XROM harmonics is not coupled to TPD or CWE but rather to a ROM-like mechanism.

\begin{figure*}[t]
	\centering
		\includegraphics[width=1.0\textwidth]{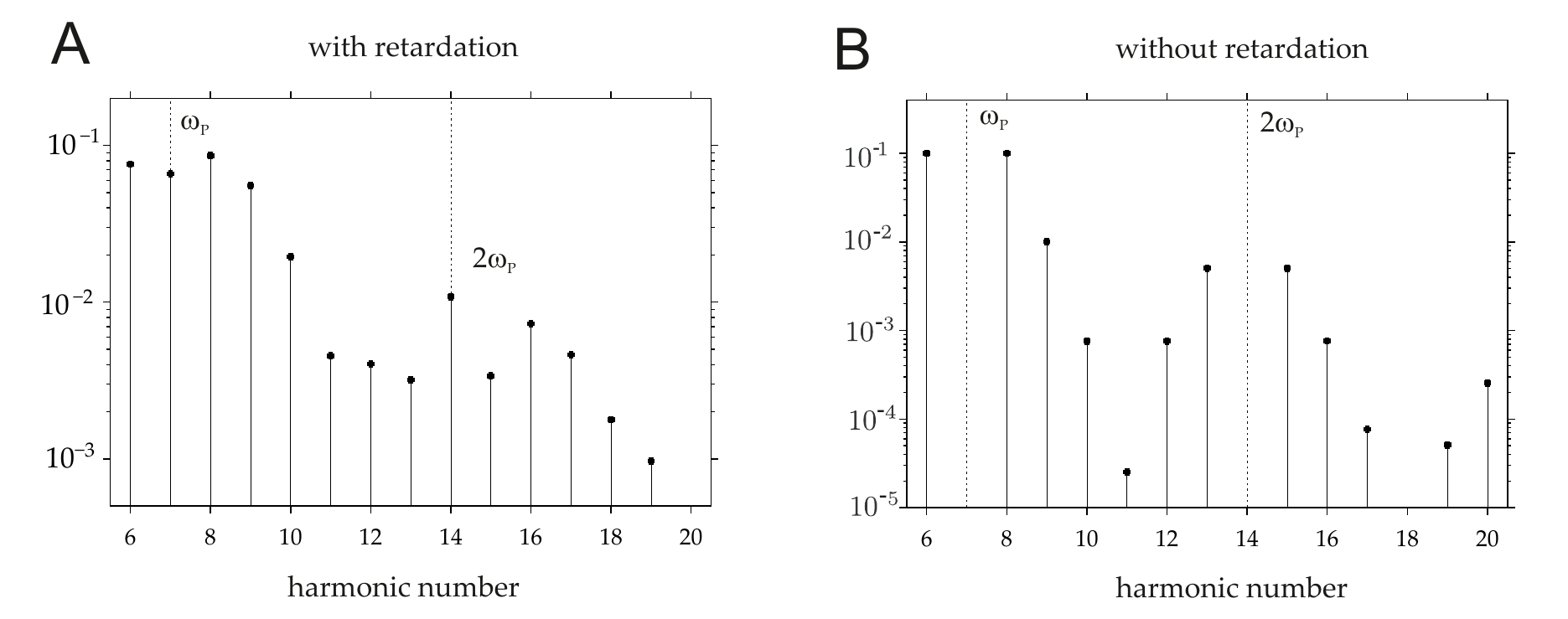}
	\caption{\textbf{Harmonic spectra from the ROM model using two oscillations at $\omega_L$ and $\omega_P$:} The parameters are taken from the PIC simulation ($\omega_P=7\omega_L$, $a_L=0.1 \, c/\omega_L$, $a_P=0.1 \, c/\omega_P$, $\varphi_L=0$, $\varphi_P=0$). \\ \textbf{(A)} When retardation $t'=t-x(t')/c$ is included in the ROM model, the harmonic amplitudes at $2\omega_P$ (14th harmonic) and $2\omega_P + 2\omega_L$ (16th harmonic) are enhanced as it is observed in the experiment. \\
\textbf{(B)} When retardation is neglected, i.e. $t'=t$,  the efficiency of the high order harmonics is strongly decreased. Moreover, the high frequency oscillation with $\omega_P$ causes enhanced sidebands at $\omega_P \pm n \omega_L$ and  $2\omega_P \pm n \omega_L$. However, neither the harmonic at $2\omega_P$ nor at $2\omega_P + 2\omega_L$ is enhanced. 
Accordingly, the relativistic effect of retardation is decisive for explaining the enhanced harmonics.}
	\label{synthesizer}
\end{figure*}

Particle-in-cell (PIC) simulations were performed to further elucidate the origin of the enhanced harmonics. For simulation parameters that closely match our experimental conditions, the temporal evolution of the electron density at the plasma surface is shown in Fig.~\ref{PIC}B and clearly reveals an oscillation not only at the optical frequency $\omega_L$ of the 400-nm laser but also a strong excitation at the plasma frequency $\omega_P$ in the XUV \cite{comment}. Although the amplitude of this oscillation $a_p=\beta^{\rm max}_p/\omega_P$ is an order of magnitude smaller than the surface oscillation at $\omega_L$, its peak velocity $\beta^{\rm max}_p \approx 0.1 c$ is comparable as seen with the chain rule. Therefore, the surface plasma oscillation at $\omega_P$ can also be expected to result in relativistic nonlinear effects in the reflected spectrum. The mechanism for the excitation of the strong plasma surface oscillation at $\omega_P$ is likely due to the jets of electrons which are expelled to the vacuum side by the laser field and which are later reinjected into the dense plasma \cite{Brunel1987, Geindre2010} (see Fig.~\ref{PIC}B). The computed reflected field analyzed by a time-windowed Fourier transform which is displayed in Fig.~\ref{PIC}A. The time integrated spectrum shows the enhanced harmonic at $2\omega_P$ on top of the familiar spectral decay of the harmonics and thus reproduces the experimental result (cf. Fig.~\ref{spectra}C). 
The strongest emission of the enhanced harmonic at $2\omega_P$ can be found in the second half of the laser pulse. A similar effect can be seen for the ROM harmonics at $\omega_P \pm \omega_L$, i.e. the sideband frequencies of the $\omega_P$ oscillation. This delayed emission of the enhanced harmonics suggests that the relativistic surface plasma mode first needs to grow during a few laser cycles. 

For analytically modeling the XROM harmonics, the reflected field is calculated in the spirit of the ROM model \cite{Tarasevitch2009}:
\begin{equation}
E_r(r)=\frac{1+\beta(t')}{1-\beta(t')} \cos(\omega_L t' + 2 k_L x(t'))
\end{equation}
with the retarded time $t'=t-x(t')/c$. The trajectory of the mirror
\begin{equation}
x(t') = a_L \cos(\omega_L t' + \varphi_L) + a_P \cos(\omega_P t' + \varphi_P)
\end{equation}
exhibits a low frequency oscillation at $\omega_L$ and, in addition to the regular ROM model, a high frequency oscillation at $\omega_P$. The oscillation amplitudes  $a_L$, $a_P$ and phases $\varphi_L$, $\varphi_P$ can be estimated using the PIC simulation (see~Fig.~\ref{PIC}B). The resulting harmonic spectra are shown in Fig.~\ref{synthesizer}. The first signature of the additional modulation to be expected are sidebands at $\omega_P \pm n\cdot\omega_L$. For parameters which are obtained from the PIC simulation, a strong enhancement spanning several harmonic orders is predicted for the harmonics around $\omega_P$. This enhancement, however, can hardly be observed in the experiment because the harmonics up to $\omega_P$ are dominated by the aforementioned CWE mechanism. More importantly, the harmonic amplitudes at $2\omega_P$ and $2\omega_P+2\omega_L$ are strongly increased in agreement with the experiment. Interestingly, the strong feature around $2\omega_P$ which is observed in the experiment and PIC simulations cannot be reproduced when retardation is neglected, see Fig.~\ref{synthesizer}B. Moreover, the harmonics' efficiency is much lower when retardation is neglected. Accordingly, retardation provides the major contribution to the nonlinearity of the ROM process and is vital for the enhancement of the XROM harmonics.

In conclusion, we found a strong enhancement of harmonic frequencies generated from a plasma surface oscillating at XUV frequencies with relativistic velocity. The enhanced harmonic generation is verified by laser plasma simulations and analytical modeling. The coherent harmonic radiation at $2\omega_P$ may be exploited in the future, for instance, as a novel diagnostic for solid density plasmas. It may further find applications in various scientific fields where intense, coherent XUV radiation of small bandwidth is needed. Examples are the investigation of plasmas in fusion-related research or in laboratory astrophysics since the radiation at $2\omega_P$ is transmitted. Other applications such as XUV spectroscopy, XUV microscopy, coherent diffraction imaging, or the seeding of free-electron lasers are evident.

%


%


\begin{thebibliography}{}
\bibitem{note}
Note that for generating harmonics at a certain photon energy, only half the harmonic order is required when SHG pulses instead of the fundamental are used. This can explain the higher efficiencies of the ROM harmonics in this frequency range when SHG pulses are used.
\bibitem{comment}
The plasma resonance has a considerable width due to damping. As it is excited by the laser frequency $\omega_L$, the dominant mode will have a frequency equal to an integer multiple of $\omega_L$.
\end{thebibliography}

\end{document}